\documentstyle[12pt]{article}
\begin{document}

\def\lsim{\mathrel{\rlap{\lower3pt\hbox{\hskip0pt$\sim$}}
    \raise1pt\hbox{$<$}}}         
\def\gsim{\mathrel{\rlap{\lower4pt\hbox{\hskip1pt$\sim$}}
    \raise1pt\hbox{$>$}}}         

\newcommand{\beq}{\begin{equation}}
\newcommand{\eeq}{\end{equation}}
\newcommand{\aver}[1]{\langle #1\rangle}

\newcommand{\La}{\overline{\Lambda}}
\newcommand{\Lam}{\Lambda_{\rm QCD}}
\newcommand{\Lv}{\Lambda_V}

\newcommand{\lam}{\lambda}

\newcommand{\lhs}{{\em lhs} }
\newcommand{\rhs}{{\em rhs} }

\newcommand{\ind}[1]{_{\begin{small}\mbox{#1}\end{small}}}
\newcommand{\hscale}{\mu\ind{hadr}}

\newcommand{\appa}{\mbox{\ae}}
\newcommand{\al}{\alpha}
\newcommand{\as}{\alpha_s}
\newcommand{\GeV}{\,\mbox{GeV}}
\newcommand{\MeV}{\,\mbox{MeV}}
\newcommand{\matel}[3]{\langle #1|#2|#3\rangle}
\newcommand{\state}[1]{|#1\rangle}
\newcommand{\ra}{\rightarrow}
\newcommand{\ve}[1]{\vec{\bf #1}}

\renewcommand{\Im}{{\rm Im}\,}

\newcommand{\eq}[1]{eq.\hspace*{.15em}(\ref{#1}) }
\newcommand{\eqs}[1]{eqs.\hspace*{.15em}(\ref{#1}) }

\newcommand{\re}[1]{Ref.~\cite{#1}}
\newcommand{\res}[1]{Refs.~\cite{#1}}

\begin{titlepage}

\renewcommand{\thefootnote}{\fnsymbol{footnote}}

\begin{flushright}
CERN-TH/96-304\\
UND-HEP-96-BIG\hspace*{.15em}$04$\\
hep-ph/9610425\\
\end{flushright}
\vspace{.8cm}
\begin{center} \LARGE
{\bf BLM-Resummation and OPE in Heavy Flavor Transitions } 
\end{center}
\vspace*{.5cm}
\begin{center}
{\Large
Nikolai Uraltsev}
\vspace*{.7cm}\\
{\normalsize
{\it TH Division, CERN, CH 1211 Geneva 23, Switzerland},\\
{\it Dept.of Physics, Univ. of Notre Dame du Lac, Notre Dame,
IN 46556, U.S.A.}\\
and\\
{\it Petersburg Nuclear Physics Institute,
Gatchina, St.Petersburg 188350, Russia}\footnote{Permanent address}
\vspace*{1.8cm} }\\
{\Large{\bf Abstract}}\vspace*{.6cm}\\
\end{center}

\noindent
An all-order resummation is performed for the effect of the running of the 
coupling $\as$ 
in the zero recoil sum rule for the axial current and for the kinetic operator
$\vec{\pi}^{\,2}$. The perturbative corrections to well-defined objects 
of OPE turn out 
to be very moderate. The renormalization of the kinetic operator is addressed.

\hfill

\begin{flushleft}
CERN--TH/96--304\\

October 1996 
\end{flushleft}
\end{titlepage}
\addtocounter{footnote}{-1}

\newpage

The recent progress in the theoretical description of heavy flavors is 
based on application of
the heavy quark expansion to QCD. Its main elements are nonrelativistic
expansion and the Operator Product Expansion (OPE) allowing a treatment  
of the strong interaction domain in model-independent systematic way. The
precision achieved in certain cases is high and requires already a careful
field-theoretic definition of effective operators beyond a simple
quantum-mechanical (QM) description. Such completely defined operators can be
introduced in different ways, however, the common feature is their scale
dependence. Likewise, the coefficient functions are always $\mu$-dependent.
The general idea of separating two 
domains and applying different theoretical tools to them was formulated long
ago by K.~Wilson \cite{wilson}
in the context of problems in statistical mechanics; in the
modern language, applied to QCD it is similar to lattice gauge theories. The
treatment of essentially Minkowskian quantities has, however, some
peculiarities. They were first considered in \cite{optical} and then in
\cite{upset}.

In this paper I apply this technique to the zero recoil sum rules
\cite{vcb,optical} to calculate one loop-corrections accounting completely for
the effect of running of the strong coupling, which can be called an extended
BLM \cite{blm} approximation, or, in short, merely BLM approximation. It is
thus a direct BLM-generalization of the perturbative calculations of
\cite{optical}. In this way the BLM-improved
perturbative evolution of the kinetic energy
operator $\bar Q (i\vec D\,)^2 Q =\vec{\pi}^{\,2}(\mu)$ and the Wilson
coefficient function $\xi_A(\mu)$ for the zero recoil sum rule for axial
current are obtained.

\section{Zero recoil sum rules}

The heavy quark sum rules and the method to calculate the relevant perturbative
effects to them have been discussed in detail in \cite{optical}; I will quote
here only necessary equations.

The first sum rule for the spatial component of the axial current at zero
recoil through terms terms $1/m_Q^2$ has the form
\beq
I_0^{(1)A}(\mu)= \xi_a(\mu)-\frac{1}{3} \frac{\mu_G^2(\mu)}{m_c^2}-
\frac{\mu_\pi^2(\mu)-\mu_G^2(\mu)}{4}\left(\frac{1}{m_c^2}
+\frac{1}{m_b^2}+\frac{2}{3m_cm_b}
\right)
\label{2}
\eeq
Here $I_0^{(1)A}$ is the zeroth moment of the first structure function $w_1$ of
a heavy hadron for the axial current:
$$
w_1^A\;=\; 2\;\frac{1}{3}\Im h_{ii}^A
$$
\beq
I_0^{(1)A}(\mu)\;=\; \frac{1}{2\pi}\, 
\int_0^\mu w_1^A(\epsilon) \,d\epsilon\;\;,\;\;\; \epsilon=
M_B-M_{D^*}-q_0\;\;.
\label{3}
\eeq
Here and in what follows the ultraviolet cutoff in the moments is assumed to be
introduced via $\theta(\epsilon-\mu)$; of course, $\mu\ll m_Q$ must hold. I use
the standard notations for the expectation values
\beq
\mu_\pi^2 = \frac{1}{2M_{H_Q}} \matel{H_Q}{\vec{\pi}^{\,2}}{H_Q}\;,\;\;\;\;
\mu_G^2 = \frac{1}{2M_{H_Q}} \matel{H_Q}{\bar Q \frac{i}{2}
\sigma_{\al\beta} G^{\al\beta} Q}{H_Q}\;.
\label{4}
\eeq
The sum rule and the lower bound for $\mu_\pi^2$ was obtained using the sum
rule for the vector structure function $w_1^V$; for simplicity of notations
below I consider the pseudoscalar ``weak'' current $J^P=\bar c i\gamma_5b$ and
the corresponding structure function $w^P$ (the result for any similar current
is the same):
\beq
I_0^{P}(\mu)\;= \; \frac{1}{2\pi}\,
\int_0^\mu w_1^P(\epsilon) \,d\epsilon\;= \;\xi^{\vec{\pi}^{\,2}}
(\mu) \,+\, \left(\frac{1}{m_c}
-\frac{1}{m_b}\right)^2 \frac{\mu_\pi^2(\mu)-\mu_G^2(\mu)}{4}\;.
\label{5}
\eeq
The Wilson coefficients $\xi$ depend on the concrete field-theoretic definition
of $\mu_\pi^2$ and $\mu_G^2$ (in what follows terms $\sim \as \Lam^2/m_Q^2$
are neglected and, therefore, $\mu_G^2$ can be considered $\mu$-independent,
$\mu_G^2\simeq 3(M_{B^*}^2-M_B^2)/4$).
At tree level one has
\beq
\xi_{A}(\mu)\; = \;1\;\;,\;\;\;\xi^{\vec{\pi}^{\,2}}(\mu)\; =\;0\;\;.
\label{6}
\eeq

The first -- and so far the only -- definition of the kinetic operator in 
field theory was given in \cite{optical}: the expectation value was defined to
have perturbatively $\xi^{\vec{\pi}^{\,2}}=0$ at any $\mu$. Thus, to calculate
$\mu_\pi^2$ in an external gluon field one must solve the Dirac (Pauli) 
equation of motion for nonrelativistic heavy quark, find the spectrum and
integrate the spectral density induced by the current $\vec\sigma 
\vec\pi$ from $0$ to
$\mu$. For an external soft field with frequencies much below $\mu$ this yields 
the classical value of $(i\vec{D}\,)^2$ \cite{optical}, therefore this is a
proper definition of the operator $\vec{\pi}^{\,2}$ in the quantum field
theory. In general it clearly depends on $\mu$. Given the definition, one is
able to calculate its $\mu$-dependence (mixing with unit operator) and the
value of $\xi_A$ without neglecting terms $\sim \mu^2/m_Q^2$ in the latter.

\section{Calculation}

The technical method was explained at length in \cite{optical,upset}; it 
reduces to considering the OPE expansion in perturbation
theory. Moreover, $\mu_G^2$ vanishes in perturbation theory to leading
order in $1/m_Q$; it will be discarded below. In general, the kinetic
operator is singled out by averaging operators over the hyperfine ($Q$-spin
related) multiplet whereas the chromomagnetic operator is projected out.

The new element here is the calculation of effects of running $\as$. In a 
one loop
Euclidean calculation it is done by replacing $\as$ by $\as(k^2)$ where $k$ is
the gluon momentum. For Minkowski quantities, in particular those that are not
obtained by analytic continuation from Euclidean space, one should use the 
dispersive
approach (it was used as long as 20 years ago and recently attracted renewed
attention; for the extensive analysis and recent discussion see 
\cite{tau,wdm}). It relies on 
using the dispersion relation for the dressed gluon propagator:
\beq
\frac{\as(k^2)}{k^2} \left(\delta_{\mu\nu}-c\frac{k_\mu k_\nu}{k^2}\right)\;=\;
-\frac{1}{\pi}\;\Im \int \;\frac{d\lam^2}{\lam^2}\;\frac{\Im
\as(-\lam^2)}{k^2+\lam^2}\; \left(\delta_{\mu\nu}-c\frac{k_\mu
k_\nu}{k^2}\right)\;.
\label{8}
\eeq
Then $(\delta_{\mu\nu}-ck_\mu k_\nu/k^2)/(k^2+\lam^2)$ is a propagator of a
gluon with mass-squared $\lambda^2$ (the longitudinal component does not
contribute in the one loop diagrams), and 
\beq
\rho\;=\;-\frac{1}{\pi}\,\Im \as(-\lam^2)
\label{8a}
\eeq
plays the role of the weight function for integrating over $\lam^2$. More
details are found in \cite{wdm} and papers mentioned there.

As the first step a calculation of all quantities of interest in the one loop
approximation with a non-zero gluon mass is needed. 
At $\lam^2=0$ the calculations have
been done in \cite{optical} and the modifications are straightforward. The
perturbative spectral densities in the \lhs of the sum rules are given by the
``elastic'' peak of $b\ra c$ transition with $\epsilon=0$ and by the transition
into the final state $c +g(\vec{k})$ with
$\epsilon=\omega(\vec{k})=\sqrt{\lam^2+\vec{k}^{\,2}}$ starting 
$\epsilon_{\rm
min}=\lam\,$. The effect of the quark recoil is of higher order in $1/m_Q$
and is discarded for $1/m_Q^2$ terms. In the sum rule for 
$\mu_\pi^2$ 
\beq
\frac{1}{2\pi}\, 
\int_0^\mu w_1^P(\epsilon) \,d\epsilon\;= \;\left(\frac{1}{2m_c}
-\frac{1}{2m_b}\right)^2 \mu_\pi^2(\mu)
\label{7a}
\eeq
the elastic transition is forbidden by parity.

The inelastic transition amplitudes given by Figs.~1 are 
modified minimally compared to
calculation of \cite{optical} but require certain care. One should distinguish
between $\omega$ and $|\vec{k}|$ and use the proper two-body phase space 
$|\vec{k}|\,d\omega$; the scalar denominators of the
propagators are $\pm 2m_Q \omega$. Finally, the massive gluon has 
three polarization propagating, instead of two for the massless one. The
following trick simplifies the calculation.

Since the tree gluon emission amplitude is transverse, the gluon propagator can
arbitrarily be redefined in the following way:
\beq
\delta_{\mu\nu}\; \ra \; \delta_{\mu\nu}\;- a_\mu k_\nu - \tilde a_\nu k_\mu
\label{10}
\eeq
Choosing $a=\tilde a = (k_0, -\vec{k})/2k_0^2 \equiv \tilde k/2k_0^2$ we arrive
at the the tensor part of the propagator 
\beq
\tilde g_{\mu\nu} = \delta_{\mu\nu} - \frac{1}{2k_0^2}(\tilde k_\mu k_\nu+
\tilde k_\nu k_\mu) = \sum_{i=1,2} e_{\mu\perp}^i e_{\nu\perp}^i +
\frac{\lam^2}{k_0^2} \, n_\mu n_\nu\;,\;\;\; n=(0, \vec{k}/|\vec{k}|)\;.
\label{11}
\eeq
In doing so we explicitly excluded the Coulomb quanta. The last term gives the
contribution of the longitudinal polarization. In this form the rest of the
calculation is simple.

For the perturbative sum rule (\ref{7a}) we get 
\beq
\frac{1}{2\pi}\,w^P(\omega;\;\lam) \, d\omega \;=\; 
\theta(\omega-\lam) \: \frac{2\as}{3\pi}
\left(\frac{1}{m_c} -\frac{1}{m_b}\right)^2 
\left(1+ \frac{1}{2}\frac{\lam^2}{\omega^2}\right)\; |\vec{k}|\, d\omega\;;
\label{12}
\eeq
the effect of the gluon mass is only kinematic besides the longitudinal
contribution $\lam^2/2\omega^2$ in the last bracket.

As in the original analysis \cite{optical} one can consider the sum rule for
$\frac{1}{3}\gamma_k\times \gamma_k$ currents:
\beq
\frac{1}{2\pi}\,
\int_0^\mu w_1^V(\epsilon) \,d\epsilon\;= \;
\frac{\mu_\pi^2(\mu)-\mu_G^2(\mu)}{12}
\left[ \left(\frac{1}{m_c}
+\frac{1}{m_b}\right)^2 + 2 \left(\frac{1}{m_c}
-\frac{1}{m_b}\right)^2 \right] + \frac{\mu_G^2(\mu)}{3m_c^2}\;.
\label{13}
\eeq
The OPE states that {\em all} soft physics is given by the \rhs; the
perturbative $\mu$-dependence of the moment is thus given by that of the
operators appearing in it. Differentiating over $\mu$ we would get the
perturbative spectral density, which thus must have the same functional form as
\eq{12} with the proper dependence on quark masses. A direct computation yields
\beq
\frac{1}{2\pi}\,
w_1^V(\omega;\;\lam) \;=\; \frac{2\as}{9\pi}\;\theta(\omega-\lam)\: 
\left[ \left(\frac{1}{m_c}
+\frac{1}{m_b}\right)^2 + 2 \left(\frac{1}{m_c}
-\frac{1}{m_b}\right)^2 \right]
\left(1+ \frac{1}{2}\frac{\lam^2}{\omega^2}\right)\; |\vec{k}|\, \;;
\label{14}
\eeq
the OPE, of course, is not violated in the one loop perturbative
calculations.\footnote{There were a few explicit checks \cite{bbz} and even 
attempts to 
disprove \cite{ns} OPE at this level in the recent years.}

An (apparently) alternative definition of $\mu_\pi^2(\mu)$ was also given in
\cite{optical,third} using the third sum rule at $\vec v \ne 0$ in the small
velocity (SV) kinematics $|\vec v\,|\ll 1$: 
\beq
\frac{\vec v^{\,2}}{3} \, \mu_\pi^2(\mu)\; = \;
\frac{\int_0^\mu \; w(\epsilon,\, v)\,
\epsilon^2\, {{\rm d}\epsilon} }{\int_0^\mu \; w(\epsilon,\, v)\,
{{\rm d}\epsilon}}\;+\;{\cal O}(\vec v^{\,4},\,1/m_Q)
\label{b1}
\eeq
where $w$ is generated by an arbitrary heavy quark weak current having a 
non-vanishing nonrelativistic limit at $\vec v = 0$. In this case the
calculation is even simpler since the soft gluon radiation factorizes and for
spacelike polarizations is merely given by the classical bremsstrahlung 
amplitude  
\beq
A_i\;\propto\; g_s\cdot \frac{v_i}{\omega}\;\times \;
\mbox{color factor}\:.
\label{b2}
\eeq
With $\epsilon=\omega$ in the SV limit one immediately arrives once again at 
the same one-loop contribution
\beq
\frac{\mu_\pi^2(\mu)}{\rm d \mu}\;= \;\frac{4}{3} \frac{\as}{\pi}
\, \theta(\mu-\lam)
\,\left(1+\frac{1}{2}\frac{\lam^2}{\mu^2} \right) \cdot 2|\vec k\,|
\;,\;\;\;\; |\vec k\,|=\sqrt{\mu^2-\lam^2}
\label{b3}
\eeq
in accord with the OPE.

A similar calculation for the zero-recoil axial current sum rule (\ref{2}) 
yields a more
cumbersome result. The elastic peak is present and results in a contribution
\beq
\frac{1}{2\pi}\,
w_{1\;\rm pert}^{A \;\rm (el)} (\epsilon)\; = \;\delta(\epsilon) \cdot
\left(1+\frac{8\as}{3\pi} r_0(\lam) \right)
\label{15}
\eeq
where $r_0(\lam)$ has been calculated in \cite{bbbsl}, eq.~(B.3). The
calculation of the continuum part was detailed above and yields
\beq
\frac{1}{2\pi}\,
w_{1\;\rm pert}^{A \;\rm (cont)} (\epsilon)=
\frac{\as}{3\pi}\theta(\omega-\lam)
\left[ 2\left(\frac{1}{m_c^2}
+\frac{1}{m_b^2} +\frac{2}{3m_c m_b} \right) 
\frac{\vec k^{\,2}}{\omega^2}
 + \left(\frac{1}{m_c}
-\frac{1}{m_b}\right)^2\frac{\lam^2}{\omega^2} 
\frac{\vec k^{\,2}}{\omega^2}
\right]|\vec{k}|\,.
\label{16}
\eeq

\subsection{Resummation of perturbative corrections}

Integrals of the spectral densities \eq{12} and \eqs{15}, (\ref{16}) are
readily calculated:
\beq
\aver{\mu_\pi^2}^{\rm pert} (\mu;\,\lam)\;= \;\frac{4\as}{3\pi}\;
\theta \,(\mu^2-\lam^2) \frac{(\mu^2-\lam^2)^{3/2}}{\mu}\;\;.
\label{20}
\eeq
Inserting this result into the \rhs of the sum rule (\ref{2}) one gets
$$
\xi_A(\mu;\,\lam)= 1+ \frac{8\as}{3\pi} r_0(\lam)
+ \frac{\as}{3\pi} \left\{
\left(\frac{1}{m_c}+\frac{1}{m_b}\right)^2
\left[ -2\lam^2 \log{\frac{\mu+
\sqrt{\mu^2-\lam^2}}{\lam}}
+2\mu\sqrt{\mu^2-\lam^2}\;-
\right.\right.
$$
\beq
\left.\left.
-\;\frac{2}{3} \frac{(\mu^2-\lam^2)^{3/2}}{\mu} 
\right] +
\frac{1}{3}
\left(\frac{1}{m_c}-\frac{1}{m_b}\right)^2
\left[\frac{(\mu^2-\lam^2)^{5/2}}{\mu^3}+\frac{(\mu^2-\lam^2)^{3/2}}{\mu}
\right]
\right\}
\cdot
\theta (\mu^2-\lam^2) \;\;.
\label{21}
\eeq
Using the explicit expression for $r_0(\lam)$ \cite{bbbsl} one finds
\beq
r_0(\lam)\;= \;\frac{3}{4}\left(\frac{m_b+m_c}{m_b-m_c}\log{\frac{m_b}{m_c}} -
\frac{8}{3} \right) -\frac{1}{8} \lam^2 \log \frac{\lam^2}{m_Q^2} 
\left(\frac{1}{m_c}+\frac{1}{m_b}\right)^2 + {\cal O}(\lam^2)\;;
\label{22}
\eeq
$\xi_A(\lam)$ does not contain nonanalytic in $\lam^2$ terms through order
$1/m_Q^2$ which shows the absence of $1/m_Q^2$ infrared (IR) renormalon
singularity in the BLM calculation. This, of course, is ensured by
OPE since the infrared contribution below $\mu$ is peeled off from 
$\xi_A$;\footnote{The residual 
nonanalytic terms $\sim \lam^3/m_Q^3$ and smaller
remain  since the way to calculate $\xi_A$ is accurate up to such terms; they
can be arbitrarily added or removed unless the operators of $D=6$ and higher
are incorporated. Their inclusion in the similar calculation would kill the
next nonanalytic terms as well.} nonanalytic terms, on the other hand, can
appear only in the infrared -- elsewhere the propagator
$1/(k^2+\lam^2)$ is analytic in $\lam^2$. The cancellation thus provides an 
independent check of the cumbersome expression for $r_0(\lam)$.

The absence of these nonanalytic terms ensures the absence of the corresponding
IR renormalon in the resummation procedure; in other words, purely 
perturbatively the IR renormalons match
in the sum rule. The concrete analysis therefore disagrees with the claims 
of \cite{ns} of a discrepancy;
basing on it the sum rules were declared erroneous in
\cite{ns,update}. The criterion as applied in those papers does not make
sense in general, and there is no point to address it here.

One also notes the absence of nonanalytic terms in 
$\aver{\mu_\pi^2}_{\rm 1\; loop}$, in spite of non-trivial operator 
mixing. It signifies
the absence of the IR renormalon in $\bar Q (i\vec D\,)^2Q$ in the BLM
approximation \cite{bbz} (see also \cite{upset}).
The reason will be touched upon later.

To calculate the BLM-type resummed result one must fix a particular form of
the strong coupling according to one's preference. Then, for an observable $A$
having the perturbative expansion through order $\as$
\beq
A(\lam^2)\;=\;1+\frac{\as}{\pi}A_1(\lam^2)
\label{23}
\eeq
one has 
\beq
A^{\rm resum}\;=\;1+ 
\int \;\frac{d\lam^2}{\lam^2}\;A_1(\lam^2)\, \rho(\lam^2)\;\;.
\label{24}
\eeq
For calculating $\xi_A$ the exact infrared behavior of the coupling is not
essential since the IR domain has been removed.

The most popular choice for ``all-order'' resummation is the literal one-loop
$\as$:
\beq
\as(k^2)\;=\;\frac{\as(Q^2)}{1+\frac{b}{4\pi} \log{\frac{k^2}{Q^2}} } \;= \;
\frac{4\pi}{b\log{\frac{k^2}{\Lv^2}} }
\label{25}
\eeq
with
\beq
\rho^{\rm BLM}(\lam^2)\;=\; \frac{4}{b}\: \left(
\Lv^2\,\delta(\lam^2+\Lv^2) \;-
\; \frac{1}{\log^2{\frac{\lam^2}{\Lv^2}}+\pi^2}\right)\;\;.
\label{25a}
\eeq
$\Lv$ is $\Lam$ in the $V$-scheme \cite{blm}, $\Lv = {\rm e}^{5/6} 
\Lam^{\overline {\rm MS}} \simeq 2.3 \Lam^{\overline {\rm MS}}$; 
here and throughout the
paper $\as$ is the $V$-scheme coupling: 
$\as(k) = \as^{\overline {\rm MS}}({\rm
e}^{-5/6}k)$ neglecting higher-order effects in the $\beta$-function.
In this case one has
\beq
A^{\rm BLM}\;=\;1+\;\frac{4}{b} \left[A_1(-\Lv^2) - 
\int \;\frac{d\lam^2}{\lam^2}\;A_1(\lam^2)\, 
\frac{1}{\log^2{\frac{\lam^2}{\Lv^2}}+\pi^2}
\right]\;\;.
\label{26}
\eeq
Using the explicit expression \eq{21} the all-order BLM result for $\xi_A$ is
obtained by simple integration.

The $\mu$-dependence of $\mu_\pi^2$ and of the coefficient function $\xi_A$ 
$$
\frac{{\rm d}\, \vec\pi^{\,2}(\mu)}{\rm d\mu}\;=\;2\,c_\pi 
(\as(\mu))\; \mu \cdot 
{\rm I}
$$
\beq
\frac{{\rm d}\, \xi_A(\mu)}{\rm d\mu}\;=\;2\,\appa_A (\as(\mu))\; \mu 
\label{28}
\eeq
in this approximation is given by integrating directly the continuum parts of 
$w^{\rm pert}$:
$$
c_\pi\;=\; \frac{4}{3}\,
\int \;\frac{d\lam^2}{\lam^2}\: \rho^{\rm BLM}(\lam^2)\;
\left(1+ \frac{1}{2}\frac{\lam^2}{\omega^2}\right)\,
\sqrt{1-\frac{\lam^2}{\mu^2}}\, \;
\theta(\mu^2-\lam^2) 
$$
$$
\appa\;=\;\frac{1}{9}\;
\int \;\frac{d\lam^2}{\lam^2}\: \rho^{\rm BLM}(\lam^2)\;
\left\{
4\,\left(\frac{1}{m_c}+\frac{1}{m_b}\right)^2
\left(1-\frac{\lam^2}{4\mu^2}\right)
\;+\right.
$$
\beq
\left.
+\;
2\,\left(\frac{1}{m_c}-\frac{1}{m_b}\right)^2
\left(1+\frac{\lam^2}{2\mu^2} - \frac{3\lam^4}{4\mu^4}\right)
\right\}\,
\sqrt{1-\frac{\lam^2}{\mu^2}} \;\, \theta(\mu^2-\lam^2)
\label{29}
\eeq
Although the integrals here formally run over the ``infrared domain'' of 
$\lam^2$ as well, they actually do not include infrared effects due to 
specific constraints on the moments of $\rho(\lam^2)$ \cite{bsg} 
with the weights in \eqs{29}
analytic below $\mu^2$.

In a similar way one could have tried to resum the perturbative series in
$\mu_\pi^2(\mu)$ itself:
\beq
\Delta_{\pi^2}^{\rm BLM}(\mu)\; \equiv \; \aver{\mu_\pi^2}_{\rm pert}(\mu)
\;=\; \frac{4}{3} \int \;\frac{d\lam^2}{\lam^2}\: \rho^{\rm BLM}(\lam^2)\;
\frac{(\mu^2-\lam^2)^{3/2}}{\mu}\:\theta(\mu^2-\lam^2)\;.
\label{30}
\eeq
However, this procedure is physically senseless, no such object can be
defined and any correction beyond BLM makes it ill-defined. It can be
convenient, however, as an {\em intermediate} result to relate the particular 
field-theoretic definition of $\vec\pi^{\,2}$ to another one if the similar
calculations in the latter are made. The absence of the IR renormalon in the
one-loop approximation allows at least formal assigning a definite value to 
$\Delta_{\pi^2}^{\rm BLM}(\mu)$. 
For this reason its anatomy will be addressed below.

\section{Analysis}

In Figs.~2 I draw the plots of $\Delta_{\pi^2}^{\rm BLM}(\mu)$; Fig.~2a shows 
the
dimensionless ratio $\tau_{\pi^2}=\Delta_{\pi^2}^{\rm BLM}(\mu)/\Lv^2$ as a
function of $\mu/\Lv$; Fig.~2b illustrates the absolute value for three
choices of $\mu$.

The value of
\beq
\eta_A(\mu)\;\equiv\;\left( \xi_A(\mu)\right)^{1/2}
\label{40}
\eeq
is shown in Fig.~3; $\eta_A(\mu)$ must be used to calculate the physical
formfactors in the OPE approach instead of $\eta_A$ of HQET in the model
estimates. Numerical evaluation was performed assuming $m_b=4.8\GeV$ and
$z=m_c/m_b=0.3$.

For reasonable values of $\mu$ and $\Lv$ natural for low-energy physics (the
corresponding analysis \cite{volmb} suggests $\Lv\simeq 300\MeV$) the
``perturbative'' BLM-piece of $\mu_\pi^2$ is clearly moderate, $\simeq 
0.2\GeV^2$, and the latter 
does not depend too strongly on the renormalization point. Nevertheless, the
BLM series, having a finite radius of convergence, is still divergent: the
radius of convergence is given by~\footnote{The notations used here 
follow \cite{ir}
where their meaning is clarified.}
\beq
\left(\frac{\as(\mu)}{\pi}\right)_{\rm max}
\;\equiv \; \frac{\al_{\rm sh}}{\pi}\;=\;
\frac{4}{\pi b}\;\;;
\label{41}
\eeq
the series is sign-varying and asymptotically behaves like 
$$
\frac{\as}{\pi} \;\frac{\pi^2}{n^{5/2}}\, \cos{\left(\frac{\pi
n}{2}-\frac{3\pi}{4}\right)}\;
\left(\frac{\pi b}{4}\right)^n \left(\frac{\as(\mu)}{\pi}\right)^n\;\;.
$$
Even at $\Lv= 300\MeV$ the scale $\mu_{\rm sh}$ where the series starts to
converge, $\as(\mu_{\rm sh})=4/b$,  constitutes $1.44\GeV$. The first ten 
terms are written
below:
$$
\Delta_{\pi^2}^{\rm BLM}(\mu) \;\simeq\; 
\frac{4\as(\mu)}{3\pi}\cdot \mu^2 \cdot
\left[1+2.88x - 2.51x^2- 44.8x^3- 6.4x^4+980x^5+924x^6\;- 
\right.
$$
\beq
\left.
-25900x^7-40900x^8+
769400x^9+ 162000x^{10}\:+
 ...\right]\;\;;\;\;\;\;\; x=\frac{\as(\mu)}{\pi}
\label{42}
\eeq
(this series is written in terms of the $V$-scheme $\as(\mu)$).
It is clear that the first BLM correction cannot represent numerically the
complete result for any $\mu$ relevant for the heavy quark expansion in charm,
as soon as the effects of the running of $\as$ in this domain is taken into
account.

The analytic properties and the expansion are most conveniently obtained using
the representation
\beq
\Delta_{\pi^2}^{\rm BLM}(\mu) \;=\; \frac{4\as(\mu)}{3\pi}\;\frac{1}{2\pi i}\,
\oint \;\frac{d\lam^2}{\lam^2}\:
\frac{(\mu^2-\lam^2)^{3/2}}{\mu}\:\frac{\as(\mu^2)}{1+\frac{b}{4\pi}
\as(\mu^2)\log{\frac{-\lam^2}{\mu^2}}}
\label{43}
\eeq
with the contour shown in Fig.~4.

The results for $\xi_A$ are shown only in the limited range $\mu < 0.8\GeV$;
irrespectively of the actual onset of duality, values of $\mu$ above $0.7\GeV$
are not appropriate for OPE in charm: the power expansion runs in powers of 
$\mu/m_{c,b}$. With the running mass $m_c(\mu) \lsim 1.15 \GeV$ one cannot
sensibly adopt larger $\mu$ if the running of $\as$ is included: in the 
BLM approximation the perturbative corrections are {\em not} small in this 
domain and the condition $\mu \ll m_c$ must be carefully respected.

Assuming $\as$ running one cannot allow larger values of $\Lam$ either; it is
enough to recall that the ``standard'' value of $\Lam^{\overline {\rm MS}} 
\simeq 250\MeV$ implies
that the usual three-loop coupling hits the Landau singularity as early as at 
the gluon momentum $\sim 900\MeV$. These arguments having a more general
practical relevance will be discussed in more detail in a separate publication.

A clear illustration is merely the impact of $1/m_Q^3$ and higher IR
renormalons in $\xi_A$; as shown below already at $\Lam^{\overline {\rm MS}}
=220\MeV$ one has
$\delta_{1/m_Q^3}(\eta_A^2) \simeq 0.05$, thus making the uncertainty 
irreducible
without an account for $1/m_c^3$ effect larger than the perturbative
corrections themselves.

The quantity $\xi_A(\mu^2;\lam)$ calculated through terms $\mu^2/m_c^2$ 
still contains
nonanalytic terms $\sim \lam^3$,  $\lam^4 \,\log{\lam^2}$ etc.:
$$
\pi \;\frac{11+5z+5z^2+11z^3}{16} \left(\frac{\lam^2}{m_c^2}\right)^{3/2}\cdot
\frac{\as}{\pi}
$$
\beq
\frac{9+2z+2z^2+2z^3+9z^4}{18} \;\frac{\lam^4}{m_c^4} 
\log{\frac{\lam^2}{m_c^2}} 
\cdot \frac{\as}{\pi}\;\;.
\label{45}
\eeq
The results shown in Fig.~3 correspond to a linear extrapolation of
$\xi_A(\mu^2;\lam)$ in $\lam^2$ from $\lam^2=0$ to $-\Lv^2$ which almost
coincides with the principal value prescription; all arbitrariness is an effect
of $\Lam^3/m^3$ and higher terms. The corresponding uncertainties (defined as the 
formal imaginary part) 
$$
\delta_{1/m_Q^3}(\eta_A^2) \;\simeq\; 0.0105\;
\left(\frac{\Lv}{300\MeV}\right)^3\;\;\;\;\;\;\;
\delta_{1/m_Q^4}(\eta_A^2)\;
\simeq \;0.0015\;
\left(\frac{\Lv}{300\MeV}\right)^4
$$
\beq
\label{46}
\eeq
are still small at $\Lv=300\MeV$ but too significant already at $\Lv=500\MeV$.

Again, the radius of convergence of the BLM series for the Wilson coefficient
(discarding in an arbitrary way $1/m_Q^3$ and higher terms) is given by
$\as^{(V)}(\mu)=4/b\simeq 0.45$ corresponding to $\overline \as(\mu) \simeq
0.29$.

\section{Once more about the kinetic operator}

Certain confusion exists in the literature about the expectation value of the
kinetic operator $\mu_\pi^2$. In the HQET a similar quantity is denoted
$-\lam_1$; they coincide on the classical level but generally differ in the
quantum field theory (QFT). However, a similar field-theoretic 
definition of $-\lam_1$ has never been given.

On one hand, $-\lam_1$ is often defined as $\mu_\pi^2(\mu)$ from which some
calculated perturbative corrections are subtracted (different in different
approaches) \cite{lig}; on the other hand, the values deduced for such
$-\lam_1$ are also claimed for $\mu_\pi^2$ \cite{lig,others} -- which
certainly cannot be true simultaneously.

The difference between possible definitions of the renormalized operator is
seen already at order-$\as$ perturbative corrections. They are given by two
diagrams in Fig.~5; both diverge quadratically and must be cut off.
A direct computation shows~\footnote{This simple calculation has been checked
in discussions 
with M.~Shifman and A.~Vainshtein in late 1994 to clarify the observation
\cite{bbz} about the disappearance of the renormalon in the kinetic operator in
one loop.} that if one averages over
directions of $k_\al$ before integrating over $k^2$, these diagrams exactly
cancel each other, see \eqs{c1}-- (\ref{c2}). Therefore, if a cutoff is 
introduced by 
any function
depending on $k^2$, the one-loop renormalization vanishes. Clearly, one does
not get renormalization in BLM resummation either. As a matter of fact, in the
Abelian theory this holds up to the order $\al^3$, and without light flavors
($m_q \lsim \mu$) in all orders in $\alpha$ (see Appendix). In the non-Abelian 
theory,
on the other hand, such an ``isotropic'' averaging merely cannot be defined 
in a gauge-invariant way beyond one loop.

The operator $\vec{\pi}^{\,2}$, however, is defined 
differently in \cite{optical}. It is easy to 
see that at the one loop level it corresponds to
introducing $\theta(\mu-|\vec k\,|)$ with no constraints on $k_0$. In such
a case the renormalization is obviously present, and is given by
$c_\pi=4\as/3\pi$ \cite{optical}.

The `Lorentz-invariant' cutoff at $k^2=\mu^2$ at first sight seems more
convenient. However, the advantage is purely technical and applies exclusively
to the one-loop perturbative calculations.\footnote{It is worth noting that
both ways to introduce a cutoff are equally Lorentz-covariant: 
$\theta(\mu^2-\vec{k}^{\,2})= \theta(\mu^2+k^2-(vk)^2)$ where $v$ is 
velocity of the heavy hadron. The difference is rather the explicit
dependence on the hadron state (its velocity). However, this distinction does
not have a general meaning since the kinetic operator itself 
$\bar Q\,\left\{-D^2+(vD)^2 \right\}Q\,$ is, in this respect, not
Lorentz-invariant either depending explicitly on the velocity $v_\alpha$.
Alternatively, the nonrelativistic heavy quark field $\tilde Q$ is not
an invariant since is constructed from the QCD field $Q$ in the non-invariant
way $\tilde Q= {\rm e\,}^{im_Q(vx)} Q$ and, thus, evolves according to a 
non-invariant equation of motion $ i\,vD\, \tilde Q=0$ involving explicit vector
$v_\alpha$. In the non-Abelian theory the attempt to define a $v$-independent
cutoff on the gluon momenta is necessary a gauge-dependent procedure.} 
It is possible to work with such 
definitions in Euclidean theories and statistical mechanics.

On the other hand, theory of heavy quarks is essentially Minkowskian, and a
`Lorentz-invariant' cutoff generally does not allow to formulate theory on the
nonperturbative level. Imposing constraints on the timelike components of
momenta leads to non-locality in time, and the standard Hamiltonian
approach is not applicable. For example, a heavy hadron state $|H_Q\rangle$
over which one wants to average $\bar Q (i\vec D\,)^2Q$ is not completely
defined. Using another language, the Euclidean amplitudes either cannot be
continued to the physical domain, or possess wrong analytic properties there.
These deep theoretical questions will be discussed in subsequent publications.

Practically, as soon as a proper particular definition of an operator is given,
one can relate its expectation value to that in other schemes. For example, 
in routine calculations in the QCD sum rules \cite{pp} the cutoff is
extrapolated to zero in the one-loop approximation. Such a value of 
$-\lam_1$ is less than $\mu_\pi^2$ by some small amount calculated in
\cite{optical}, therefore, strictly speaking, in a number of practical 
applications
$\mu_\pi^2$ must be taken even {\em larger} than estimates quoted in
\cite{pp}. On the other hand, for purely technical reason a definition with
the covariant cutoff was applied in the calculation of the semileptonic 
widths in
\cite{upset}, therefore the literal numbers of \cite{pp} were used there,
which are somewhat smaller than $\mu_\pi^2$ discussed in \cite{optical} and in 
the present paper. The numerical analysis above easily allows one to use a more
convenient calculation scheme in each particular case. It is worth emphasizing
again \cite{optical} that subtracting the perturbative piece from the
renormalized operators does not make sense by itself when nonperturbative
corrections are addressed, and the result is usable only as long
as the amount subtracted from the expectation value of properly defined
operator is explicitly specified.

The advantage of the ``physical'' definition of $\mu_\pi^2$ is that it allows
to derive an important lower bound
\beq
\mu_\pi^2\;>\;\mu_G^2
\label{68}
\eeq
and still leads to moderate -- and well-controlled -- perturbative corrections
in observable quantities. In any case, so far no sensible 
alternative definition has been given in the literature; in principle, of
course, it can be done.

It is important to note that the relation between {\em properly} given QFT
definitions of an operator is done perturbatively and {\em excludes} the IR
domain. For example, if there is no renormalon in the perturbative series for
it 
in a certain approximation in one scheme, the same holds in any other scheme
as well. This explains the absence of the renormalon in the BLM series for
$\mu_\pi^2(\mu)$ noted in Sect.~2.1: with the cutoff over $k^2$ all terms merely
vanish.

It does not contradict the presence of mixing:
\beq
\frac{{\rm d} \mu_\pi^2(\mu)}{{\rm d} \mu^2}\;=\;c_\pi (\as(\mu))\;\;.
\label{60}
\eeq 
Even in the large-$n_f$ approximation $c_\pi$ is a whole series in $\as(\mu)$,
see \eq{29}. It is easy to understand this fact: calculating
$\mu_\pi^2(\mu)$ perturbatively  one integrates over the `cylinder' domain
$$
-\infty \;<\;k_0\;<\;\infty\;\;,\;\;\;|\vec k\,|< \mu\;.
$$
On the other hand, the integral over the spherical domain  $k^2<\mu^2$
vanishes. For this reason the integral can be reduced to the domain
$$
k_0\;>\; \sqrt{\mu^2-\vec k^{\,2}}\;,\;\;\; |\vec k\,|< \mu
$$
(see Fig.~6) where $k^2>\mu^2$ always holds and the coupling remains small.

It means that in \eq{60} the coefficient $c_{\pi}$ is given by an effective 
coupling that never grows to infinity at finite $\mu$ in the BLM approximation.
These subtleties were missed in \cite{invis} where mixing \eq{60} was
equated to the presence of renormalons.

The situation is completely different in the case of the pole mass: the leading
IR contribution $\delta m_Q \sim \mu$ comes from $k_0 \lsim \frac{\vec
k^{\,2}}{m_Q} \lsim \frac{\mu^2}{m_Q} \simeq 0\,$ \cite{pole}, 
therefore $k^2=
-\vec k^{\,2}$ and the
corresponding coupling is always $\as(\mu)$ in the BLM approximation.

\section{Conclusions}

All-order BLM effects are calculated for the zero recoil sum rules within
proper OPE approach. Contrary to existing claims they have quite moderate
impact. In particular, for the available field-theoretic definition of the
kinetic operator $\vec{\pi}^{\,2}$ its expectation value obeys the inequality
\cite{vcb,optical}
$$
\mu_\pi^2(\mu)\;>\;\mu_G^2
$$
for any normalization point $\mu$; the normalization-point dependence is very
moderate.

The conclusion differs from a recent paper \cite{lig} 
where the first nontrivial
BLM terms ${\cal O}(b\as^2)$ were addressed; those represent the first term of
the expansion of the full function calculated in the present paper. It is seen 
that the series for this functions are still divergent and the calculated terms
already grow in magnitude; in this case the estimates based on the first term
${\cal O}(b\as^2)$ 
are numerically misleading. The analysis shows that for the scale governed by
charm effects of running of the coupling -- if addressed at
all~\footnote{The phenomenological relevance of such theoretical improvement 
is far
from obvious; it will be discussed elsewhere.} -- must include the whole 
resummation, made in the OPE-consistent way.

Moreover, regarding the perturbative corrections to the sum rule for the 
zero-recoil formfactor $F_{D^*}$ (the axial-current sum rule (\ref{2})), 
the calculation of ${\cal O}(b\as^2(\mu))$ terms in \cite{lig} missed 
the similar contribution from $\eta_A^2$; considering that calculation 
as the (second-order) perturbative correction to the sum rules is thus 
erroneous. It is only the
whole set of corrections that eliminates the infrared domain from the
perturbative factor. As a result of this omission the major part of the
corrections came from the domain near and below the Landau singularity and
therefore seems irrelevant.

It is worth mentioning that the smearing procedure considered in \cite{lig}
practically does not improve the convergence: the critical value of 
$\as$ for the weight $\mu^{2n}/(\mu^{2n}+\epsilon^{2n})$
decreases by only a factor of $1-1/(2n)$ for the price of increasing 
the actual 
cutoff scale while the room for it is severely limited by the small value of
$m_c$.

The value of $\eta_A(\mu) = \left(\xi_A(\mu)\right)^{1/2}$ at a representative
choice $\mu\simeq 0.5\GeV$, $\Lam^{(V)}=300\MeV$ is 
\begin{eqnarray*}
\eta_A(\mu)\; = & 1  &  \;\;\;\mbox{ tree level}\\
\eta_A(\mu)\; \simeq & 0.975 &   \;\;\;\mbox{ one loop}\\
\eta_A(\mu)\; \simeq & 0.99  &  \;\;\;\mbox{ all-order BLM}
\end{eqnarray*}
Clearly, the perturbative corrections including BLM effects are very moderate
and differ minimally from the estimate $0.98$ used in the original analysis
\cite{vcb,optical}. The difference falls well below the effect of $1/m_c^3$
corrections not addressed so far.

The nonperturbative corrections, on the other hand, do have a grave numerical
impact on the model analysis \cite{update,othe} (see also \cite{ns}) where it
was postulated to use ill-defined $\eta_A^{\rm (HQET)}(=\eta_A(0)\,)$ according
to the routine practice of HQET \cite{neubpr}. In all later publications 
the perturbative
factor was used about $0.95$ which is not supported by the analysis. Moreover,
such an approach results in double-counting of the soft domain contributions.
\vspace*{0.25cm}

\noindent
{\bf Acknowledgements:} \hspace{.1em} 
I am grateful to I.~Bigi, M.~Shifman and
A.~Vainshtein for collaboration; the results and, in particular,
approach presented here were discussed with them in detail.
I thank V.~Braun, G.~Martinelli and C.~Sachrajda for their interest, and 
P.~Ball for invaluable help in the calculational aspects.
The presented calculations were accomplished during my stay at CERN; it is a 
pleasure to acknowledge the hospitality and the creative atmosphere of 
the Theory Division.
This work was supported in part by NSF under the grant
number PHY 92-13313.\vspace*{.9cm}

\noindent
{\Large\bf Appendix} \vspace*{.5cm}
\renewcommand{\theequation}{A.\arabic{equation}}
\setcounter{equation}{0}

\noindent
At one loop the renormalization of $\mu_\pi^2$ is given by Figs.~5. The dark
box is the operator $\bar{Q} (i\vec{D})^2 Q\, (0)\,$. The
corresponding Feynman integrals are
\beq
I_a\;=\;\int\frac{{\rm} d^4k}{(2\pi)^4i}\; \frac{1}{k^2}\, \frac{1}{k_0^2}
\, \vec k^{\,2} \qquad \qquad \qquad \qquad
I_b\;=\;-\,\int\frac{{\rm} d^4k}{(2\pi)^4i}\; \frac{3}{k^2}\;\;.
\label{c1}
\eeq
Performing the Wick rotation $k_0=ik_4$ and averaging the factor $\vec
k^{\,2}/k_0^2$  yields
\beq
\int\;{\rm} d^4k\; \frac{k^2-k_4^2}{(k_4+i0)^2}\, \delta(k^2-p^2)\;=\;
\;-3\:\int\;{\rm} d^4k\; \delta(k^2-p^2)\;\;.
\label{c2}
\eeq
Thus, with such a cutoff $I_a+I_b=0$. If, to preserve the QM interpretation,
one first integrates over $k_0$ from $-\infty$ to $\infty\,$, one obviously has
$3I_a=-I_b\,.$

Let us consider higher orders in QED without light flavors. The Lagrangian
is 
\beq
{\cal L}\;= \;\int\; {\rm d}^4 x \left[\bar Q \,iD_0 \,Q \;-\; m_Q\,\bar Q \,Q
\right](x)
\;+\; {\cal L}_\gamma
\label{c3}
\eeq
where the (Euclidean) Lagrangian of the Abelian gauge theory with the isotropic
cutoff at $k^2=\mu^2$ ($\mu \ll m_Q$) is defined as 
\beq
{\cal L}_\gamma\,= \,\frac{1}{2}\,\int\: 
\frac{{\rm} d^4k}{(2\pi)^4} \: A_\mu(k) A^*_\nu(k)\, (k^2
\delta_{\mu\nu}-k_\mu k_\nu)\: \theta(\mu^2-k^2)\,+\, 
\mbox{gauge fixing terms.} 
\label{c4}
\eeq
The dynamic degrees of freedom are $A_\mu(k)$ with $k^2< \mu^2$. Note
that, formally, the gauge invariance is lost due to the constraint on $A_\mu$;
the current is still conserved, however. In this theory $A_{1,2,3}$ are
sterile; only $A_0$ interacts with $Q$. Thus in
higher orders one has only the Coulomb-exchange dressing of two one-loop 
diagrams Figs.~5, illustrated in Figs.~7a and 7b, respectively. It will be 
shown below that both are renormalized multiplicatively by
the same factor (which, as a matter of fact, equals unity) and thus the 
overall renormalization of the kinetic operator vanishes if the
renormalization scheme is such that two first-order graphs in Fig.~5 cancel.

This follows from the exact factorization in the emission of the Coulomb 
quanta. Since
the gauge field self-interaction is absent (generally, it is a
cutoff-dependent fact), the perturbation theory is fully 
time-ordered in the heavy quark limit. It is convenient to classify the
perturbative diagrams by the number $n$ of Coulomb quanta spotted at $t=0$.
Consider first the tree-level emission of $n$ quanta from the initial state
(terms $e^k$ for the amplitude with $k=n$).
The corresponding sum over possible permutations is given by 
\beq
\sum_{\rm permutations}
\frac{1}{\omega_1}\cdot\frac{1}{\omega_1+\omega_2} \cdot
\frac{1}{\omega_1+\omega_2+\omega_3} \cdot \,...\;=\; \prod_i\;
\frac{1}{\omega_i}\;.
\label{c5}
\eeq
All virtual emissions cancel between the wavefunction and vertex
renormalizations due to the QED Ward identity  ($Z_1=Z_2$), and possible
formfactors are absent in the limit $\mu/m_Q \ra 0$. Thus, in this static limit
the emission amplitude for $n$ photons is {\rm exactly}
$e^n\prod\frac{1}{\omega_i}$ to all orders. This property is nothing but the 
quantum decomposition of the classical Coulomb field of a static source.

The Feynman integral for the sum of the diagrams Fig.~7a with $n$ Coulomb
quanta at $t=0$ is then given by
$$
I_a^{(n)}\;=\;\frac{(4\pi\alpha)^n}{n!}\;\int\;
\prod_{l=1}^{n}\;
\frac{{\rm d}^4k_l}{(2\pi)^4i}\:
\frac{1}{k_l^2\omega_l^2} \cdot 
\left(\sum_l \vec{k}_l\right)^2\;=
$$
\beq
=\;
\frac{(4\pi\alpha)^n}{(n-1)!} \;\int\;\frac{{\rm d}^4k_n}{(2\pi)^4i}\,
\frac{\vec{k}_n^{\,2}}{k_n^2\omega_n^2} \;\;\int\; 
\prod_{l=1}^{n-1}\;\frac{{\rm d}^4k_l}{(2\pi)^4i}\:
\frac{1}{k_l^2\omega_l^2}
\label{c6}
\eeq
whereas, with $k_n$ being the momentum of the spacelike-polarized photon in 
the blob, the expression for the diagrams Fig.~7b with $n\!-\!1$ Coulomb quanta 
is clearly 
\beq
I_b^{(n)}\;=\;-\,
\frac{(4\pi\alpha)^n}{(n-1)!} \;\int\;\frac{{\rm d}^4k_n}{(2\pi)^4i}\,
\frac{3}{k_n^2} \;\;\int\; 
\prod_{l=1}^{n-1}\;\frac{{\rm d}^4k_l}{(2\pi)^4i}\:\frac{1}{k_l^2\omega_l^2}
\;\;.
\label{c7}
\eeq
Thus the diagrams Fig.~7a and 7b cancel order by order. Note that the common
factors in eqs.~(\ref{c6}) and (\ref{c7}) are merely the 
vertex correction $Z_1$ for the static charged particle to the given order,
which coincides with the wavefunction renormalization $Z_2$; it 
cancels the diagrams with $n=0$. 
Therefore the one-loop 
value of $\mu_\pi^2$ simply is not renormalized by higher orders if the
regularization yields a non-vanishing value in order $\alpha$. 
This is also rather obvious starting from the definition \eq{b1}.

Since the result above does not depend on a particular form of the gauge 
propagator and only requires it to be an invariant function of $k^2$, the
insertion of fermion loops does not destroy the cancellation as long as
effective multi-photon vertices do not appear. Due to Furry's theorem it
occurs only at order $\alpha^3$ when light fermions are present; examples
are shown in Figs.~7c, 7d.

A similar cancellation is not expected to hold in the non-Abelian case 
already at two loops \cite{upset} where the difference with the Abelian case
first emerges. However, the isotropic cutoff by itself cannot be defined in a
gauge-invariant way; for example, the simplest Pauli-Villars regularization
violates color current conservation, and the renormalization, instead, would
depend on the gauge. There is a more complicated general method to introduce a
similar regularization in the invariant way \cite{reg}; the presence of mixing
there is, however, the most general feature in two loops.

\newpage
\vspace*{-1.0cm} 

\noindent 
{\large\bf Figure Captions}\vspace*{0.8cm}

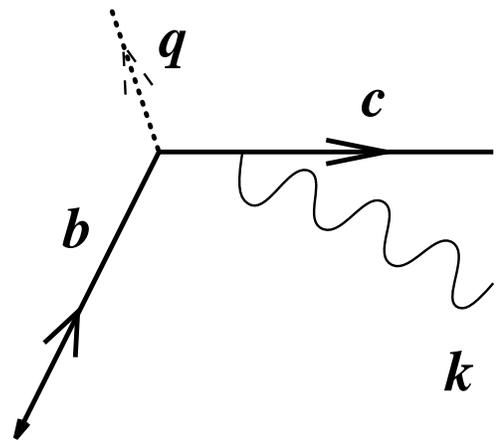
\noindent {\bf Fig.~1:} \hspace*{.1em} Perturbative diagrams leading to 
nontrivial inelastic structure functions.\\

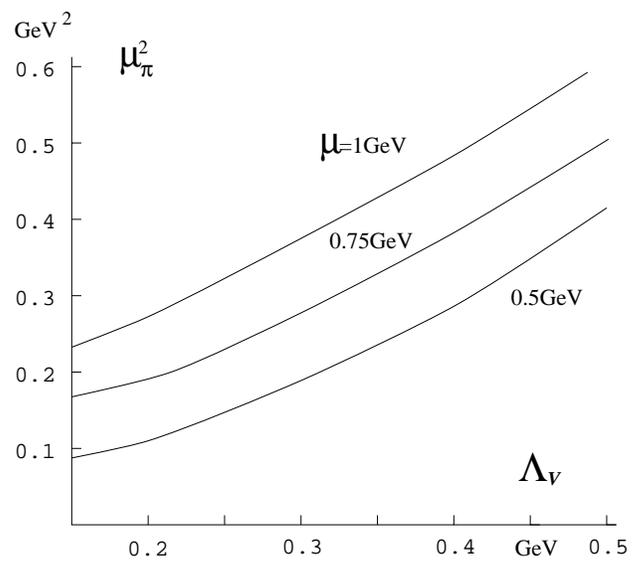
\noindent {\bf Fig.~2:}\hspace*{.1em} The perturbative part of the kinetic
operator in the resummed BLM approximation, $\Delta_{\pi^2}^{\rm BLM}(\mu)$.\\
{\bf a):} The dimensionless ratio $\tau_{\pi^2}=\Delta_{\pi^2}^{\rm
BLM}(\mu)/\Lv^2$ as a function of the dimensionless scale parameter
$\mu/\Lv$.\\
{\bf b):} The absolute value of $\Delta_{\pi^2}^{\rm BLM}(\mu)$ as a function of
$\Lam^{(V)}$ for $\mu=0.5\GeV$, $\mu=0.75\GeV$ and $\mu=1\GeV$.\\

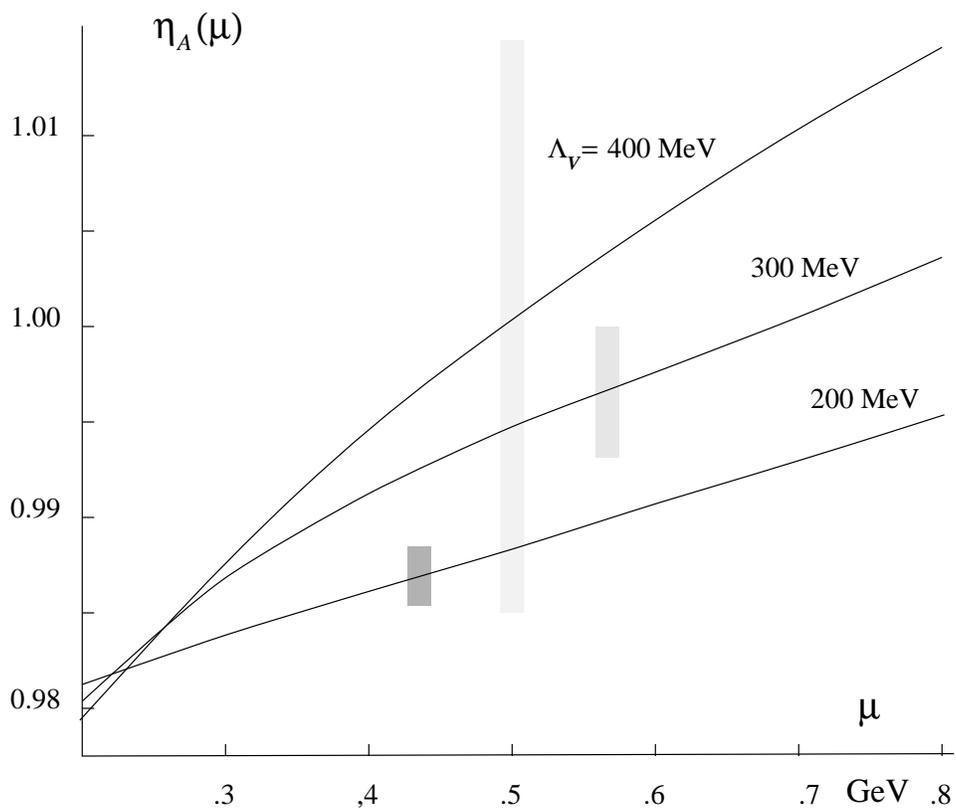
\noindent {\bf Fig.~3:}\hspace*{.1em} The value of the Wilson coefficient
$\eta_A(\mu)=\left( \xi_A(\mu)\right)^{1/2}$ for $\Lam^{(V)}=200\MeV$,
$\Lam^{(V)}=300\MeV$ and $\Lam^{(V)}=400\MeV$. The value of $\eta_A(\mu)$ must
be used in the QCD-based calculations of the exclusive zero-recoil $B\ra D^*$ 
formfactor when nonperturbative effects are addressed. The shaded bars show the
purely perturbative uncertainty irreducible without the OPE account for 
$1/m_c^3$ and $1/m_c^4$ effects; they thus represent the lower bound for 
the corresponding actual nonperturbative corrections.\\

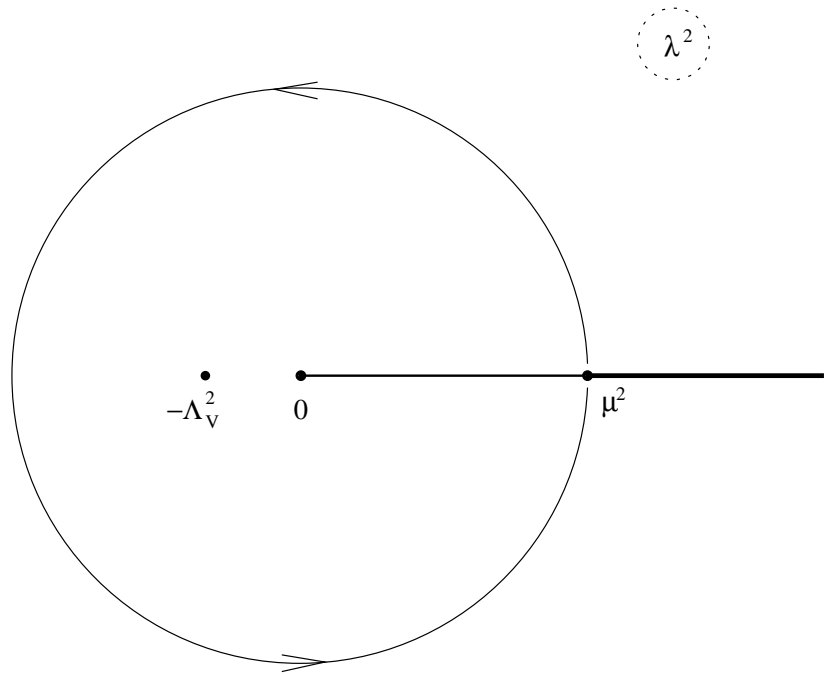
\noindent {\bf Fig.~4:}\hspace*{.1em} The contour of integration over $\lam^2$
allowing a straightforward perturbative expansion of $\Delta_{\pi^2}^{\rm
BLM}(\mu)\,$.\\

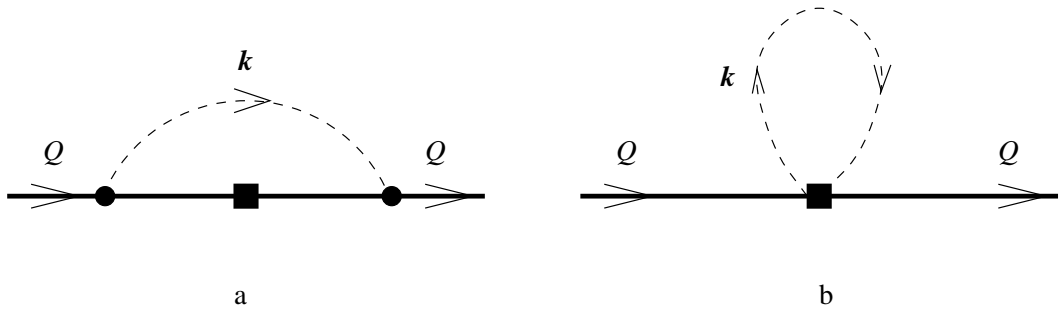
\noindent {\bf Fig.~5:}\hspace*{.1em} Feynman diagrams contributing to the
one-loop renormalization of the operator $\bar Q\, (i\vec D\,)^2\,Q\,$. Dashed 
line is the gauge boson (gluon); dark box represents the operator. In the
diagram a) only the Coulomb quanta ($A_0$) propagation contributes whereas in b)
the gluon is spacelike.\\

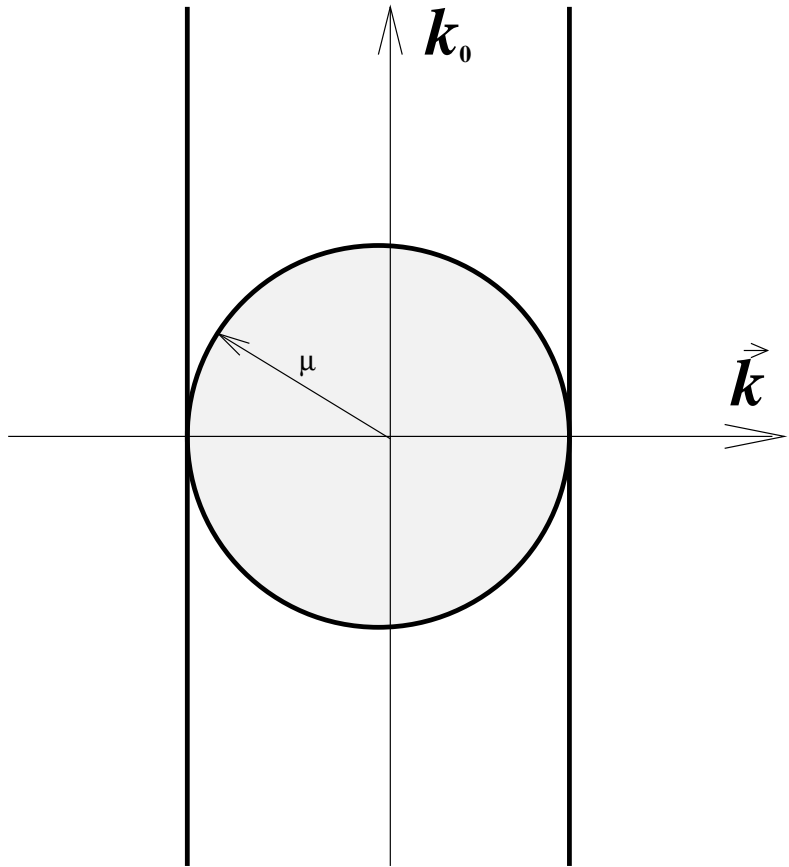
\noindent {\bf Fig.~6:}\hspace*{.1em} The domain of integration over the 
gluon momentum
$k$; the cylinder $-\mu <|\vec{k}\,|< \mu$ gives the renormalization of
$\mu_\pi^2$ in one loop. The integral over the shaded disk (sphere)
vanishes.\\

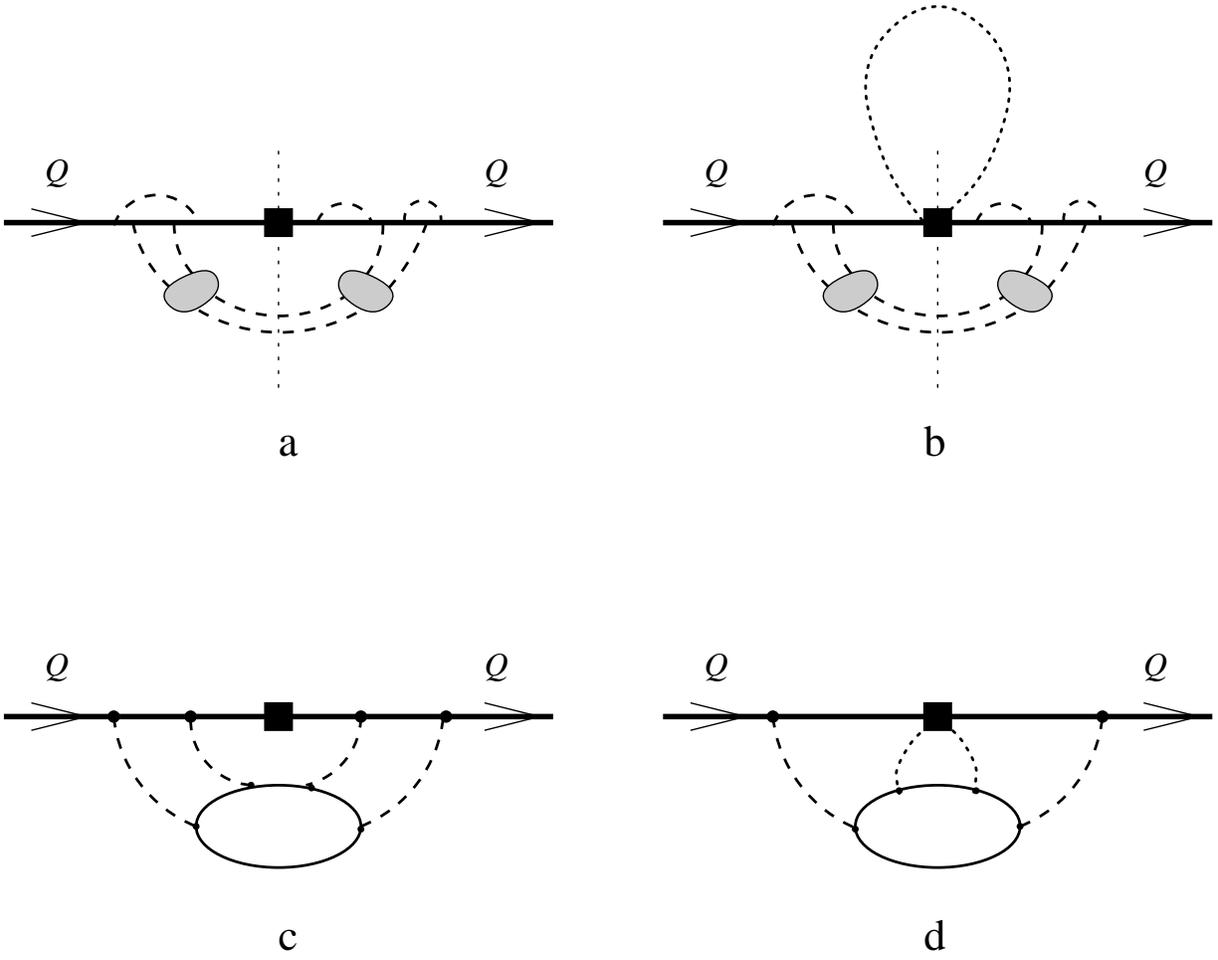
\noindent {\bf Fig.~7:}\hspace*{.1em} Higher-order renormalization of the
kinetic operator in the Abelian theory.\\ {\bf a}, {\bf b}: without light 
flavors ($n=2$).\\
 {\bf c}, {\bf d}: presence of light flavors 
induces 
photon self-interaction in the order $\alpha^2$. \\
Dashed photon
lines denote Coulomb quanta, dotted lines are for the spacelike polarizations.
Dark box is $\bar Q\, (i\vec D\,)^2\,Q(0)\,$. Shaded blobs show the 
possibility to twist photon lines.

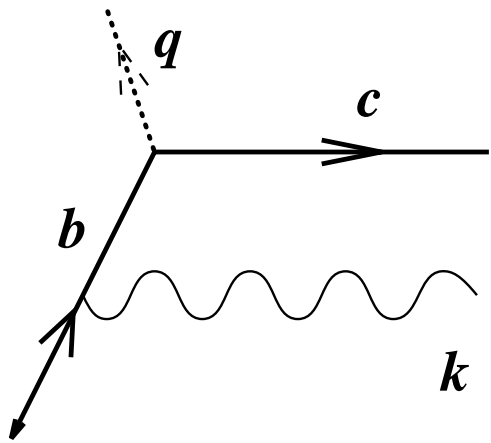
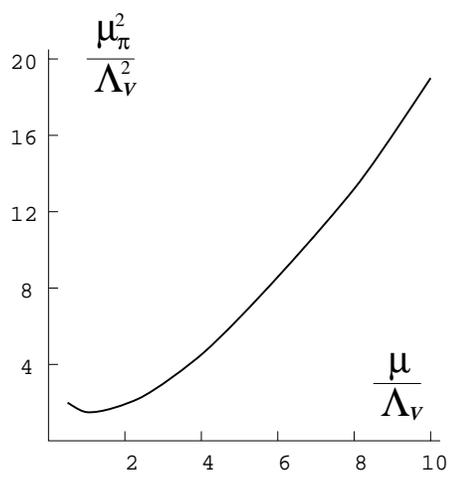
\end{document}